\documentclass[aps, prl,twocolumn, showpacs,superscriptaddress,10pt,floatfix]{revtex4-2}

\usepackage{graphicx}
\usepackage{dcolumn}
\usepackage{bm}
\usepackage{hyperref}
\hypersetup{
    unicode=false,          
    pdftoolbar=false,        
    pdfmenubar=true,        
    pdffitwindow=false,     
    pdfstartview={FitH},    
    pdfkeywords={Quantum batteries} {Ergotropy} {Micromasers}{Steady State}, 
    pdfnewwindow=true,      
    colorlinks=true,       
    linkcolor=black,          
    citecolor=blue,        
    filecolor=magenta,      
    urlcolor=blue           
}

\usepackage{xcolor}
\usepackage{mathrsfs}
\usepackage{amsfonts}
\usepackage{amsmath}
\usepackage{amssymb}
\usepackage{subfigure}
\usepackage{physics}
\usepackage[page]{appendix}

\begin{document}


\title{Micromasers as Quantum Batteries}

\author{Vahid Shaghaghi}
\email{vshaghaghi@uninsubria.it}
\affiliation{Center for Nonlinear and Complex Systems, Dipartimento di Scienza e Alta
Tecnologia, Universit\`{a} degli Studi dell’Insubria, via Valleggio 11, 22100 Como, Italy}
\affiliation{
Istituto Nazionale di Fisica Nucleare, Sezione di Milano, via Celoria 16, 20133 Milano, Italy}
\affiliation{Center for Theoretical Physics of Complex Systems, Institute for Basic Science (IBS), Daejeon - 34126, Korea}

\author{Varinder Singh}
\email{varinder@ibs.re.kr}
\affiliation{Center for Theoretical Physics of Complex Systems, Institute for Basic Science (IBS), Daejeon - 34126, Korea}

\author{Giuliano Benenti}
\email{giuliano.benenti@uninsubria.it}
\affiliation{Center for Nonlinear and Complex Systems, Dipartimento di Scienza e Alta
Tecnologia, Universit\`{a} degli Studi dell’Insubria, via Valleggio 11, 22100 Como, Italy}
\affiliation{
Istituto Nazionale di Fisica Nucleare, Sezione di Milano, via Celoria 16, 20133 Milano, Italy}
\affiliation{NEST, Istituto Nanoscienze-CNR, Piazza S. Silvestro 12, 56127 Pisa, Italy}

\author{Dario Rosa}
\email{dario\_rosa@ibs.re.kr}
\affiliation{Center for Theoretical Physics of Complex Systems, Institute for Basic Science (IBS), Daejeon - 34126, Korea}
\affiliation{Basic Science Program, Korea University of Science and Technology (UST), Daejeon - 34113, Korea}

\date{\today}

\begin{abstract}

We show that a micromaser is an excellent model of quantum battery.  
A highly excited, pure, and effectively steady state of the cavity mode, charged by coherent qubits, can be achieved, also in the ultrastrong coupling regime of field-matter interaction. Stability of these appealing features against loss of coherence of the qubits and the effect of counter-rotating terms in the interaction Hamiltonian are also discussed.
\end{abstract}

\maketitle


\emph{Introduction.---} 
Given  current trend to miniaturization of technology, with devices operating at dimensions of the order of the nanoscale, quantum effects must be taken into account. 
The possibility of using genuine quantum effects to improve performances of such devices compared to their classical counterparts --- generically dubbed as \textit{quantum advantage} --- has in turn led to enormous efforts and research,  to find all tasks where a quantum advantage can be obtained~\cite{qcbook,Arute2019,Zhong2020-tr,Campaioli2017,Rossini2020}.

Any quantum machine needs energy to operate.
Therefore, in parallel with developments on building nanodevices, topic of \textit{quantum batteries} (see~\cite{Campaioli2019,Bhattacharjee2021}, and references therein) has emerged and flourished in the last few years.
A quantum battery is a quantum mechanical system suitable to store energy in some highly excited states, to be released on demand.
Starting from the seminal work of Alicki and Fannes, \cite{alicki_fannes_original}, several figures of merit, to characterize the performance of a quantum battery, have been introduced and studied.
These include energy storage \cite{Andolina2018,Zhang2019,Caravelli2020,Quach2020,Crescente2020}, work extraction \cite{alicki_fannes_original,Hovhannisyan2013}, fluctuations properties \cite{Friis2018,Rossini2019,Santos2019,Rosa2020} and charging power \cite{Binder2015,Campaioli2017,Le2018, Ferraro2018, Andolina2019, Crescente2020b, Ghosh2020, Rossini2020, GarciaPintos2020, hamma_comment, hamma_reply_to_comment,
Zakavati2021, Ghosh2021, Seah2021, mondal2022periodically, kanti2022quantum}, to name some of them. 

In parallel with such theoretical developments, proposals to concretely build quantum batteries have emerged, including models realized by spin chains \cite{Le2018} and qubit systems charged by electromagnetic fields \cite{Ferraro2018}. 
However, with some very recent notable exceptions \cite{Quach2022}, achievements on actual experimental realizations have been so far quite limited.
As for any quantum information protocol, the search for high-speed operations is vital. Such possibility is naturally offered by 
circuit quantum electrodynamics (QED),
one of the most promising platforms for quantum hardware,
in which one can address the so-called ultrastrong coupling (USC) regime of light-matter interaction, where the qubit-cavity interaction energy becomes comparable, 
or can even exceed the bare frequencies of the uncoupled systems \cite{FornDiaz2019,Kockum2019}.

In this paper 
we revisit, with an eye towards features persisting in the USC limit, a very well-established model, which has been extensively implemented and studied at the experimental level: the \textit{micromaser}~\cite{meschede1985one, Filipowicz_number_states, slosser_PRL, slosser_dissipation, slosser_PRA, lekien1993generation, meystre_book}, where a stream of qubits (two-level atoms in cavity QED~\cite{Walther2006}) 
sequentially interact with a cavity mode with a high-quality factor. That is, 
the radiation decay time is much larger than the characteristic time of the qubit–field interaction, and the overall evolution is to a 
good approximation coherent.
%
%
We will show that micromasers can be regarded as \textit{excellent} models of quantum batteries.
In particular, they display excellent performances in terms of \textit{charging temporal stability} 
and \textit{ergotropy}~\cite{Allahverdyan2004}. 

By charging temporal stability we mean the ability of  battery to keep, under time evolution and after an initial transient time, a stable value of its mean energy  \cite{Friis2018,Rossini2020,Santos2019,Rosa2020}.
In many models of quantum batteries, after an initial regime, in which  energy grows since  battery is evolving towards excited states, time evolution of  mean energy undergoes temporal fluctuations.
These fluctuations are of course unwanted, since they require a very high precision in controlling the charging time, in order to reach the target value of mean energy.
We will numerically show that a micromaser can be quickly charged to reach an \textit{almost steady state}, whose energy is controlled by the physical parameters defining the model.
By almost steady state we mean a state which, although metastable, is characterized by very long lifetime.
An almost steady state solves the problem of temporal stability, since it is constant for very long times.

By ergotropy it is meant the amount of energy that can be extracted from a battery via unitary operations \footnote{The notion of ergotropy assumes a perfect knowledge of the battery state as well as the possibility of performing any possible unitary operations. In many-body contexts, both these assumptions are questionable. In consequence, the notion of ergotropy has been critically re-analyzed and an improved definition, more suitable for many-body systems, has been proposed in \cite{safranek2022appear}.}.
Often, part of energy is locked in correlations and  cannot be used.
For example, when the battery state is a mixture it cannot be transformed into the ground state by means of unitary transformations.
Therefore not all its energy can be extracted.
On the opposite side, a pure state has ergotropy which equals its mean energy.
We will show that for a micromaser the almost steady states mentioned above are approximately \textit{pure} and therefore,
in principle, almost all their energy 
can be reversibly extracted.
This is a very surprising feature, since dynamics in micromasers involves a trace over qubit degrees of freedom after each interaction.

The possibility of building pure states (steady or not) in micromasers has been discussed in previous studies \cite{Filipowicz_number_states, slosser_PRL, slosser_dissipation, slosser_PRA, lekien1993generation}.
However, they relied on several assumptions: very weak coupling between qubits and field and/or \textit{highly fine-tuned} values of these couplings.
All these assumptions are unwanted in a quantum battery: weak coupling implies slow charging and consequently low values of charging power, while fine-tuned values require very high precision in building the battery, thus making it hard to realize.
Our numerical results show that all these assumptions can be relaxed to a large extent: the battery reaches an \textit{ almost pure steady state} even when entering in  USC and without fine-tuning.

The combination of above features promotes micromasers to  status of very robust and reliables quantum batteries, thus making them as very promising models for experimental realizations.

\emph{Model and figures of merit.---} We consider a quantum battery made up of a quantized electromagnetic field in a cavity, modeled as a harmonic oscillator.
It is initially prepared in its ground state $\ket{0}$.
The charging protocol is realized via a stream of two-level systems (qubits) which sequentially interact with the battery, thus realizing a micromaser.
The initial state of each qubit is 
\begin{align}
    \label{eq:initial_state_qubit}
    \rho_q &= q \ket{g} \bra{g} + (1 - q) \ket{e} \bra{e} + \nonumber \\
    &  + c \sqrt{q (1 - q)} \left( \ket{e} \bra{g} + \ket{g} \bra{e}\right) \, ,
\end{align}
where $\ket{g}$ and $\ket{e}$ are ground and  excited state of the qubit, respectively. 
Parameters $q$ and $c$ control the degrees of population inversion and coherence, respectively.
The evolution of the system is described, in interaction picture, by the Hamiltonian \cite{scully_book, meystre_book}
\begin{equation}
    \label{eq:Rabi_hamiltonian_interaction}
    \hat{H}_I = g \left( \hat{a} \hat \sigma_+ + \hat{a}^\dagger \hat \sigma_- + e^{i 2 \omega t}\hat{a}^\dagger \hat \sigma_+ + e^{-i 2 \omega t} \hat{a} \hat \sigma_- \right) \, .
\end{equation}
where $\omega$ is the frequency of both qubit and  field (we are considering the \textit{resonant} case in which they are equal, \cite{meystre_book}); $\hat{a}^\dagger$, $\hat{a}$ are  creation/annihilation operators for the field and  $\hat{\sigma}_+$, $\hat{\sigma}_-$ are  raising and lowering operators for the qubit (respectively). 
Finally, $g$ is the coupling constant for  interaction between  qubit and  field and we take units such that $\hbar = 1$.
It is customary to name the first two terms in Eq.~\eqref{eq:Rabi_hamiltonian_interaction} as \textit{rotating terms} and the last two terms as \textit{counter-rotating terms} \cite{meystre_book, scully_book}.

From Eq.~\eqref{eq:Rabi_hamiltonian_interaction} we derive the time evolution operator, $\hat U_I(g, \, \omega, \, \tau)$, to be
\begin{align}
    \label{eq:time_evolution_op_general}
    &\hat U_I(g, \, \omega, \, \tau) \equiv \mathcal{T} \exp{ - i \int_0^\tau \hat{H}_I(t) \mathrm d t },
\end{align}
where $\mathcal{T}$ and $\tau$ are the time ordering operator and the interaction time between a single incoming qubit and the battery, respectively. 
We fix, without loosing generality, $\tau = 1$.
Accordingly, $\hat U_I(g, \, \omega) \equiv \hat U_I(g, \, \omega, \, \tau = 1)$ is the time evolution operator.

Denoting with $\rho_B(k)$ the battery state after having interacted with $k$ qubits, we denote by $\rho(k)$ the product state of $\rho_B(k)$ and the incoming $(k + 1)$-th qubit, $\textit{i.e.}$ $\rho(k) = \rho_B(k) \otimes \rho_q$.
The system is then evolved by means of $\hat U_I(g, \, \omega)$, and $\rho_B(k + 1)$ is obtained as follows:
\begin{equation}
    \label{eq:one_step_evolution}
    \rho_B(k + 1) = \Tr_q\left(\hat U_I(g, \, \omega) \rho(k)\hat U_I^\dagger(g, \, \omega) \right) \, ,
\end{equation}
where $\Tr_q$ is the trace over qubit degrees of freedom.
Although the battery is initially in a pure state $\ket{0}$, by tracing out qubit degrees of freedom it becomes mixed, in general.
More concretely, one expects that the purity of $\rho_B$, $\mathcal{P}(k)$, defined as $\mathcal{P}(k) \equiv \mathrm{Tr}\left( \rho_B^2(k) \right)$, should decrease from $1$ (which is the value for pure states) before the first collision to almost vanish after many collisions.

The energy stored after $k$ collisions amounts to $E(k) \equiv \Tr \left(\hat{H}_B \rho_B(k) \right)$, where $\hat{H}_B \equiv \omega \hat{a}^\dagger \hat a $ is the battery Hamiltonian and trace is taken over battery degrees of freedom.
We now turn to characterize the charging performance of the battery by measuring $E(k)$ and $\mathcal{P}(k)$.

\emph{Numerical results.---}
The time evolution operator, Eq.~\eqref{eq:time_evolution_op_general}, gets enormously simplified when counter-rotating terms in Eq.~\eqref{eq:Rabi_hamiltonian_interaction} can be omitted.
As it is known, \cite{scully_book, meystre_book}, counter-rotating terms can be omitted when \textit{weak-coupling} condition $\frac{g}{\omega}\ll1 $ is met.
However, to build a battery which is fast in charging, weak-coupling \textit{is not} the most relevant limit. 
Hence, it will be desirable to consider the case in which $g$ and $\omega$ are comparable, such that $0.1<\frac{g}{\omega}<1 $.
This case can be treated, still without the need of considering counter-rotating terms, by performing a simultaneous modulation of 
the frequencies of field and qubit,
as first observed in \cite{ultrastrong_JC}.
We now assume that such a modulation has been performed and consider the celebrated Jaynes-Cummings (JC) operator \cite{JC_original}, 
\begin{equation}
\label{eq:jaynes_cummings_operator}
\hat{U}_I(g) = e^{- i g \left( \hat{a} \hat \sigma_+ + \hat{a}^\dagger \hat \sigma_- \right)}.
\end{equation}

\begin{figure*}[htp]
   \centering
   \subfigure[]{\label{fig:USC_combined}\includegraphics[scale=0.52]{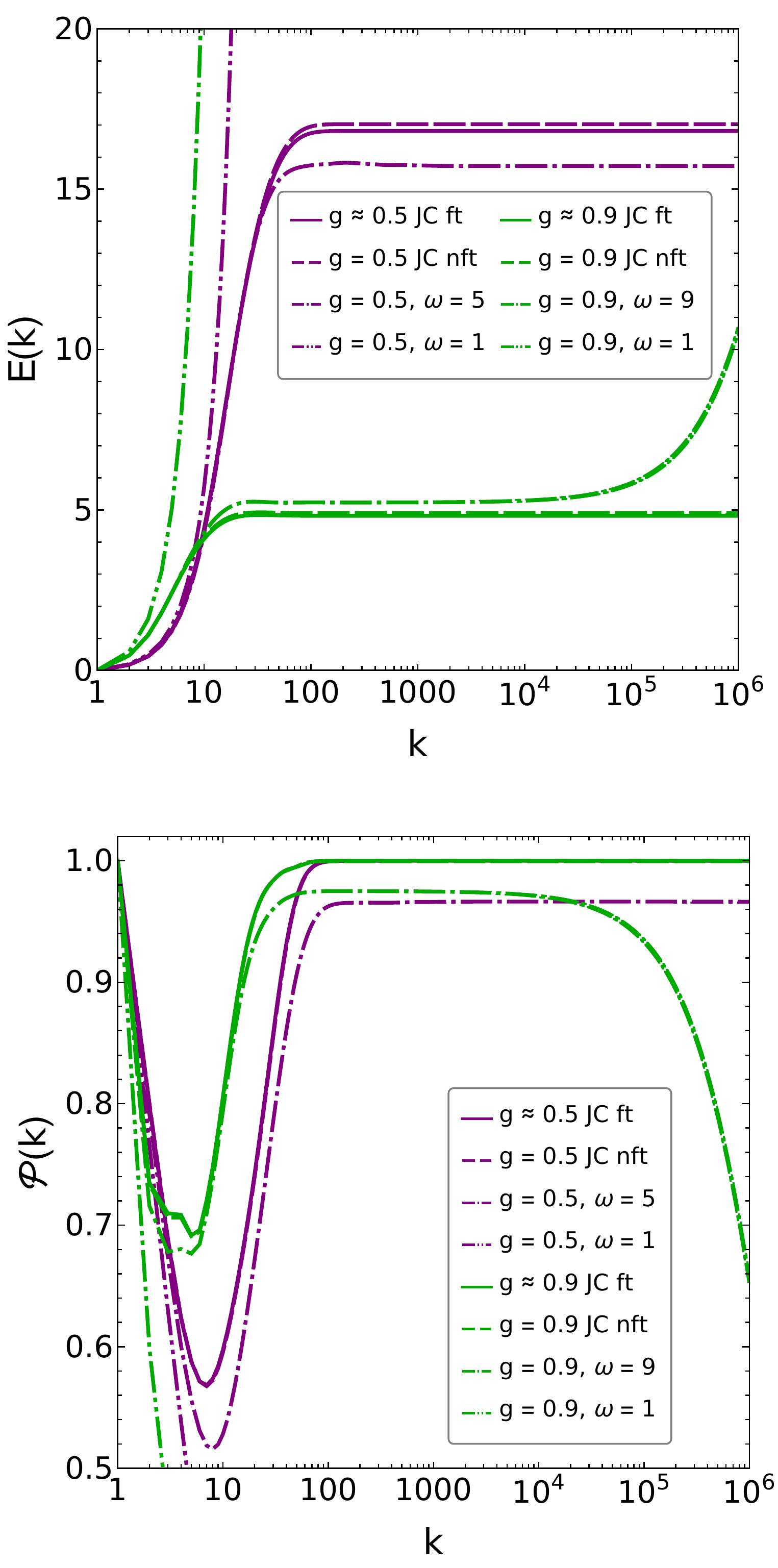}}\hspace{50pt}
   \subfigure[]{\label{fig:plotDensity}\includegraphics[scale=0.52]{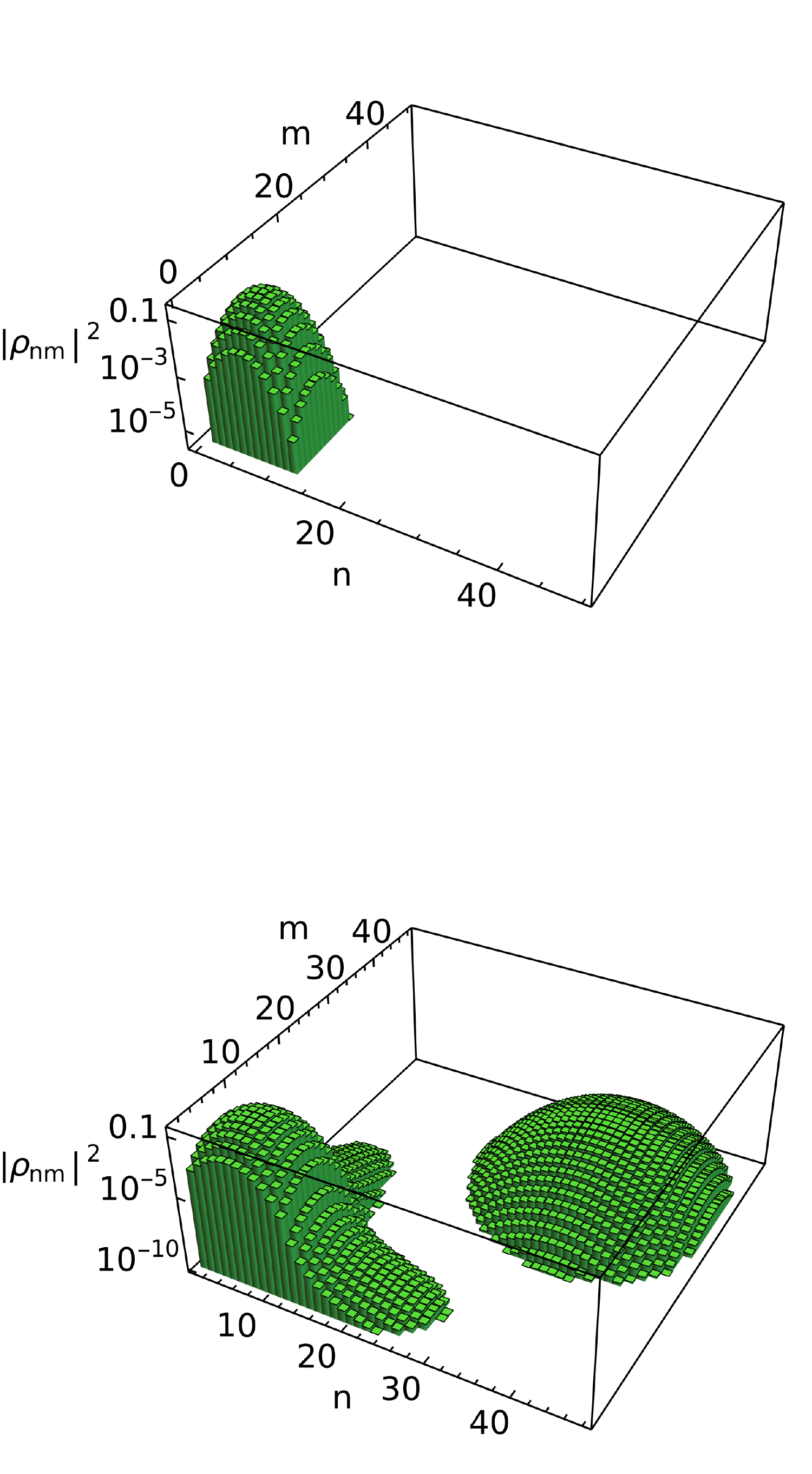}}
   \caption{Performance of the micromaser quantum battery. FIG.~\ref{fig:USC_combined}: energy stored (upper panel) and purity (lower panel) computed for $g = 0.5, \, 0.9$ and $q = 0.25$ in  JC model (denoted ``JC nft'' in legend) as well as in presence of finite values of $\omega$ and non-vanishing contributions from counter-rotating terms.
   As a comparison, we report also the charging performances for JC batteries having the closest fine-tuned values for $q = 1$, $g = \frac{\pi}{\sqrt{39}} \approx 0.503$ and $g = \frac{\pi}{\sqrt{12}} \approx 0.906$ (denoted ``JC ft'' in legend). 
   For purity, it is essentially impossible distinguishing fine-tuned from non fine-tuned dynamics.
   FIG.~\ref{fig:plotDensity}: absolute values of density matrix elements, $\abs{\rho_{nm}}^2$, after $10^6$ collisions, for fine-tuned case $g = \frac{\pi}{\sqrt{12}} \approx 0.906$ (upper panel) and for non fine-tuned case $g = 0.9$ (lower panel). 
   }
\end{figure*}

Micromasers driven by JC operator have been heavily studied in literature, starting from \cite{Filipowicz_number_states, slosser_PRL, slosser_dissipation}, both numerically and analytically.
The most important feature of the JC operator, in our context, is the existence of \textit{trapping chambers}.
It was analytically shown in \cite{Filipowicz_number_states, slosser_PRL, slosser_PRA} that when $c = 1$ and, more crucially, when $g$ is \textit{fine-tuned} to satisfy
\begin{equation}
    \label{eq:fine-tuning_condition}
    g = \frac{Q}{\sqrt{m}} \pi, \qquad Q, m \in \mathbb{N},
\end{equation}
the system reaches a steady state given by a macroscopic superposition of number states, $\ket{\psi} = \sum_n f_n \ket{n}$, involving at most $m$ number states,  with $f_n$ satisfying a recursion relation.
From this property, $m$ defines the corresponding trapping chamber, since it bounds the largest energy eigenstate entering in the expansion of the steady state.
On the other hand, to characterize a quantum battery it is important to consider both the cases in which $g$ \textit{is not} fine-tuned and to estimate the number of collisions necessary to reach such a state.

Given any value $g$, clearly it can be written as 
\begin{equation}
    \label{eq:non_fine-tuned_def}
    g = \frac{Q}{\sqrt{m + \epsilon}} \pi, \qquad Q, m \in \mathbb{N}, \quad -0.5 < \epsilon \leq 0.5 ,
\end{equation}
\textit{i.e.} it is approximated by a certain fine-tuned value, with $\epsilon$ quantifying the error in considering such an approximation.
Moreover,  by increasing $Q$ and correspondingly $m$ $\epsilon$ can be made arbitrarily small, thus suggesting that, by taking $Q$ and $m$ very large, an almost fine-tuned dynamics could be eventually reached.
Such a picture, however, does not give any information on the number of collisions necessary to reach such an almost fine-tuned regime and, more important, on the behavior of the state during evolution.
In particular, it does not provide any information of the role played on dynamics by other approximate trapping chambers, having $Q$ and $m$ small.
These questions are crucial when dealing with a quantum battery.

To this end, we have numerically computed the micromaser time evolution, controlled by Eq.~\eqref{eq:one_step_evolution}, for two non fine-tuned values of $g$, as well as for their \textit{closest fine-tuned counterparts having Q = 1}, \textit{i.e.} for $g = \pi/\sqrt{m}$.
Results are reported in Fig.~\ref{fig:USC_combined}.
For both fine-tuned and non fine-tuned values, system reaches, after $\approx 30$ collisions, an effective steady value of $E(k)$, which remains constant up to $10^6$ collisions.
This state is essentially pure, as demonstrated by  $\mathcal{P}(k)$.
While these results are non surprising when $g$ is fine-tuned, and they 
are in agreement with 
analytical results~\cite{Filipowicz_number_states, slosser_PRL, slosser_PRA},
it is remarkable that they hold in non fine-tuned cases as well.
In particular, 
when $g = 0.5$ and $Q = 1$,  
$\epsilon \approx 0.48$
is almost maximal.
Nevertheless, battery reaches an effective steady state which, for practical purposes, is stable and pure.
These are  wanted properties for a model of quantum battery, which our results show to be present up to, at least, $10^6$ collisions.
To better investigate the properties of these states, with and without fine-tuning of $g$, we have studied the density matrix, $\rho_B$, after $10^6$ collisions.
Results are reported in Fig.~\ref{fig:plotDensity}.
As expected, when $g$ is fine-tuned all non-vanishing elements of the density matrix are strictly confined in the first trapping chamber. 
What is non-trivial and interesting is the behavior when $g$ is not fine-tuned: we see that the vast majority of non-vanishing elements are still confined in the first trap.
However, some non-vanishing elements are actually escaping and a bubble is forming around matrix element $(30,\,30)$.

This result is confirming that in the presence of just partial trapping, dynamics turns out to be extremely slow. 
For many collisions the first trap, \textit{i.e.} the case with $Q = 1$, is enough to fully control the dynamics. Furthermore, to get a reliable battery, it is important that battery charging does not depend too strongly on fine-tuned values of $g$. Fig. 1(b) (lower panel) shows that in  case of non-fine-tuned $g$ there are alternate bubbles in the system density matrix; however Fig. 1(a) confirms that for practical purposes first trapping state in the density matrix is sufficient as it is a long-lived trapping state, up to $10^6$ collisions. 
We also checked (not shown) that  results are quite stable for deviations away from  $c = 1$: for values of $c$ not too far from $1$, the battery still reaches a stable dynamics characterized by a high degree of purity, $\mathcal P \gtrapprox  c$.

\emph{Stability with counter-rotating terms.---} Although by performing high-amplitude, low-frequency modulations \cite{ultrastrong_JC}, the validity of the JC approximation can be pushed well beyond the standard weak-coupling limit ($g/\omega\ll 1$), it is important to investigate the stability of the physical features just described in the presence of the counter-rotating terms in Eq.~\eqref{eq:Rabi_hamiltonian_interaction}.  
Hence, without performing any modulations, we have studied the time evolution of the battery by means of the full time evolution operator $\hat U_I(g, \, \omega)$ in Eq.~\eqref{eq:time_evolution_op_general}. 
Importance of counter-rotating terms becomes more prominent 
when entering in USC regime, $\frac g\omega \sim 0.1$.
Accordingly, we have checked that counter-rotating terms do not significantly affects dynamics up to $\frac g\omega \sim 0.05$. 
More interestingly, we have studied $E(k)$ and $\mathcal{P}(k)$ for $\frac{g}{\omega} = 0.1$ and   larger values of $g$, \textit{i.e.} in  USC regime.
Results are reported in Fig.~\ref{fig:USC_combined}.
We see that for $\frac{g}{\omega} \geq 0.5$  all features described for  JC evolution are lost: the system does not reach any stability with energy increasing indefinitely and purity steadily decreasing.
The same is not true and turns out to be more interesting for $\frac{g}{\omega} = 0.1$.
In this case, dynamics is different for $g = 0.5$ and $0.9$, thus showing that the actual behavior is controlled not only by $g/\omega$ but also by $g$ itself.
In particular, for $g = 0.5$ we see that trapping dynamics is still effective after $10^6$ collisions, although both $E(k)$ and $\mathcal{P}(k)$ are negatively affected by counter-rotating terms.
On the other hand, when $g = 0.9$ we see that stability is preserved only up to $\approx 10^4$  collisions.
All in all, our results show that trapping properties of the JC operator are rather robust, even in USC limit and without any modulation.

\emph{Conclusions and outlook.---} We have numerically shown that a micromaser, charged by means of coherent qubits, can be thought as an excellent model of quantum battery. 
The system reaches an effectively steady state which is dynamically stable. More crucially, such state is essentially \textit{pure}, which means that all of its energy can be extracted, in principle, via unitary operations.
Finally, we have shown that our results are quite stable, even when entering in  USC regime, and they are \textit{generic}, \textit{i.e.}, they do not require fine-tuning.
Combining all our results, we conclude that a micromaser, charged in presence of an external modulation and at USC, behaves as a very reliable model of quantum battery, even when modulation is not perfect and counter-rotating terms cannot be fully ignored. 
Micromasers have been extensively studied in literature, even at experimental level \cite{weidinger1999trapping}.
Hence, we think our results show an explicit and promising model of quantum battery.

Most of  recent theoretical developments have been tailored towards systems involving \textit{many} batteries and instances where collective quantum effects improve performance, \textit{i.e.} examples of quantum advantage.
Micromasers are single body systems but they can be combined in architectures to build many-body setups, \cite{greentree2006quantum, hartmann2006strongly}.
Hence, our results show that they are promising building blocks of many-body quantum batteries, an aspect that we plan to explore in  near future.

On a more theoretical ground, our results show that, for practical purposes, pure steady states can be found in  micromaser at USC and without any fine-tuning.
It would be extremely interesting to better investigate and characterize these 
states and to find analytic arguments supporting their existence and predicting their lifetime.

\emph{Acknowledgments.---} We thank Matteo Carrega and Dominik \v{S}afr\'{a}nek for discussions and collaboration on related projects.
Vahid Shaghaghi, Varinder Singh and Dario Rosa acknowledge support by the Institute for Basic Science in Korea (IBS-R024-D1).

\end{document}